\documentclass[
    ,final            % use final for the camera ready runs
  ]
  {aipproc}

\layoutstyle{8x11double}

\begin{document}

\title{Supernova Explosions and the Birth of Neutron Stars}


\author{H.-Thomas Janka}{
  address={Max Planck Institute for Astrophysics,
  Karl-Schwarzschild-Str.~1, D-85741 Garching, Germany}
}

\author{Andreas Marek}{
  address={Max Planck Institute for Astrophysics, 
  Karl-Schwarzschild-Str.~1, D-85741 Garching, Germany}
}

\author{Bernhard M\"uller}{
  address={Max Planck Institute for Astrophysics, 
  Karl-Schwarzschild-Str.~1, D-85741 Garching, Germany}
}

\author{Leonhard Scheck}{
  address={Max Planck Institute for Astrophysics,
  Karl-Schwarzschild-Str.~1, D-85741 Garching, Germany}
}

\begin{abstract}
We report here on recent progress in understanding the 
birth conditions of neutron stars and the way how supernovae
explode. More sophisticated numerical models have
led to the discovery of new phenomena in the supernova
core, for example a generic hydrodynamic instability of the
stagnant supernova shock against low-mode nonradial deformation 
and the excitation of gravity-wave activity in the surface and
core of the nascent neutron star.
Both can have supportive or decisive influence on the inauguration
of the explosion, the former by improving the conditions for energy
deposition by neutrino heating in the postshock gas, the latter by
supplying the developing blast with a flux of acoustic power
that adds to the energy transfer by neutrinos. While recent
two-dimensional models suggest that the neutrino-driven mechanism
may be viable for stars from $\sim$8$\,M_\odot$ to at least
15$\,M_\odot$, acoustic energy input has been advocated as an
alternative if neutrino heating fails. Magnetohydrodynamic
effects constitute another way to trigger 
explosions in connection with the collapse of sufficiently
rapidly rotating stellar cores, perhaps linked to the birth 
of magnetars. The global
explosion asymmetries seen in the recent simulations offer an
explanation of even the highest measured kick velocities of 
young neutron stars.
\end{abstract}

%%%%%%%%%%%%%%%%%%%%%%%%%%%%%%%%%%%%%%%%%%%%%%%%%%%%%%%%%%%%%%%%%%%
%%
%% The below \maketitle command inserts the actual front matter data.
%% It has to follow the above declarations.
%%
%%%%%%%%%%%%%%%%%%%%%%%%%%%

\maketitle

%%%%%%%%%%%%%%%%%%%%%%%%%%%%%%%%%%%%%%%%%%%%
%% MAINMATTER
%%
%%%%%%%%%%%%%%%%%%%%%%%%%%%%%%%%%%%%%%%%%%%%%%%%%%%%%%%%%%%%%%%%%%%%%%%%%%%%
%% Headings:
%%
%% The aipproc supports three heading levels, i.e., \section,
%%	\subsection, and \subsubsection.
%%%%%%%%%%%%%%%%%%%%%%%%%%%%%%%%%%%%%%%%%%%%%%%%%%%%%%%%%%%%%%%%%%%%%%%%%%%%
%% Cross-references:
%%
%% Page numbers (\pageref) and headings can NOT be referenced in the class,
%% since before being produced, no page numbers are determined.
%%
%% Tables, figures, and equeations can be referenced by using the LaTex
%% 	commands \label and \ref. For references to equation numbers, \eqref
%%	can be used, which will print "(1)" (while \ref will result in "1").
%%
%%%%%%%%%%%%%%%%%%%%%%%%%%%%%%%%%%%%%%%%%%%%%%%%%%%%%%%%%%%%%%%%%%%%%%%%%%%%
%% Lists: 
%%
%% Standard "itemize", "enumerate", etc. list environments are supported.
%%%%%%%%%%%%%%%%%%%%%%%%%%%%%%%%%%%%%%%%%%%%%%%%%%%%%%%%%%%%%%%%%%%%%%%%%%%%
%% Urls:
%%
%% \url{} command is provided for documenting URLs.
%%%%%%%%%%%%%%%%%%%%%%%%%%%%%%%%%%%%%%%%%%%%

\section{Introduction}

Improved numerical tools and the increasing power of modern
supercomputers have brought considerable progress in modeling
stellar core collapse in the past years. It is possible now to
simulate the complex physical processes in the deep interior
of supernovae with unprecedented sophistication and detailedness.

It has become clear meanwhile that the explosions of massive stars 
are a generically multi-dimensional phenomenon. This insight was
fostered by the fact that spherically symmetric (1D) simulations,
which became available with a fully energy 
dependent solution of the Boltzmann transport problem for
the neutrinos only recently (see~\cite{janka_liebendoerfer.2005} 
for an overview and comparison of different numerical approaches)
confirmed and solidified older 1D results of the 1980's and
1990's, namely that explosions in the 1D models could not be 
obtained, neither by the prompt bounce-shock
nor by the delayed neutrino-heating mechanism, 
at least not for progenitor stars of more
than 10$\,M_\odot$ and on a timescale of roughly one second
after core bounce~\cite{janka_liebendoerferetal.2003,janka_ref:buras-II,janka_thompson.2003,janka_sumiyoshi.2005}. Moreover, the latest
generation of multi-dimensional simulations has provided 
evidence for a variety of routes that can lead to explosions 
when nonradial phenomena are accounted for. These routes seem
to depend on the properties and conditions in the progenitor 
stars like their mass and structure and the amount
of angular momentum in their core.

In the following we will briefly review these recent developments
in the multi-dimensional modeling of stellar core collapse and 
explosion, and we will critically discuss the status of the present
simulations.

A better understanding of the explosion mechanism of core-collapse
supernovae is not only important for interpreting the observable
properties of the blast, for predicting gravitational-wave and 
neutrino signals, and for determining the conditions of nucleosynthesis
processes that occur during the explosion. It is also and in
particular essential for establishing the link between the progenitor
stars and their compact remnants, thus answering questions like that 
of the mass distribution of neutron stars and of the stellar mass limit
for black hole formation, which may happen either directly during the 
core collapse or by later massive fallback when the disrupted star does 
not become completely unbound during the explosion. So far, estimates
for such scenarios have been made only on the basis of still rather
crude self-consistent explosion simulations~\cite{janka_fryer.1999} or 
by invoking assumptions about the mass cut and energy in models with
piston-driven artificial explosions (e.g., \cite{janka_woosley.2007}).

\section{Brief historical excursion}

Due to the huge gravitational binding energy liberated in neutrinos,
which carry away hundred
times more energy than needed for the explosion, these
particles have long been speculated to be the driving agent of the
stellar explosion. Colgate and White~\cite{janka_ref:colgate} in a seminal
paper in 1966 not only proposed gravitational binding energy to
be the primary energy source of core-collapse supernovae, but also
that the intense flux of escaping neutrinos transfers the energy
from the imploding core to the ejected stellar mantle. Nearly twenty years
later, Bethe and Wilson~\cite{janka_ref:bethe} were the first who
described in detail the way how this might happen, interpeting thereby
the physics that played a role in hydrodynamic simulations performed
by Wilson. They concluded that electron neutrino and antineutrino 
absorptions on the free neutrons and protons that are abundantly present
in the shock-dissociated matter behind the stalled accretion shock in the 
supernova core, are the primary agents of the energy transfer. 

These pioneering computer simulations of the so-called delayed neutrino-driven
explosion mechanism were still conducted in spherical symmetry.
The mechanism turned out to be successful only when the neutron
star was assumed to become a more luminous neutrino source by mixing
instabilities accelerating the energy transport out of its dense interior.
The thus enhanced neutrino emission led to stronger neutrino heating in
the overlying layers of the exploding star. Theoretical studies and
multi-dimensional computer models, however, suggest that convection 
and mixing instabilities inside
the neutron star do not have the necessary big effect (see, e.g.\
\cite{janka_bruenn.1996,janka_ref:buras-II,janka_dessart.2006,janka_burrows.2007}).  
Instead, the first multi-dimensional simulations, which became 
available only in the mid 1990's, demonstrated
that the neutrino-heated layer around the forming neutron star is
unstable to vigorous convective overturn~\cite{janka_herant.1994,janka_burrows.1995,janka_janka.1996,janka_fryer.1999,janka_fryer.2002,janka_fryer.2004}. 
This can raise the efficiency 
of the neutrino energy deposition and thus can have a supportive
effect on the supernova explosion. The first such two-dimensional
(i.e.\ axisymmetric) and three-dimensional simulations, however,
suffered from a severe drawback: the physics of the
neutrino transport and of neutrino-matter interactions, which is
essential for discussing the power input to the explosion,
is so complex that it could be treated only in a grossly simplified
way. In the best models at that time this was done 
by the so-called ``grey diffusion approximation''. This means that
the energy-dependence of the neutrino interactions (the cross 
sections of the most important neutrino processes typically
scale with the squared neutrino energy) was ignored and
replaced by a ``grey'' (spectrally averaged) description. Moreover,
the spatial propagation was approximated by assuming that neutrinos
diffuse through the dense neutron star medium until they decouple
and stream away from a chosen position, usually from a layer somewhat
outside of the ``neutrinosphere'', close to the surface of the compact 
remnant. The historical development of these theoretical studies of 
the supernova explosion mechanism is resumed in a recent 
review~\cite{janka_ref:janka}.

\begin{figure}[htp!]
\tabcolsep=1mm
\begin{tabular}{lcr}
  \includegraphics[width=.33\textwidth]{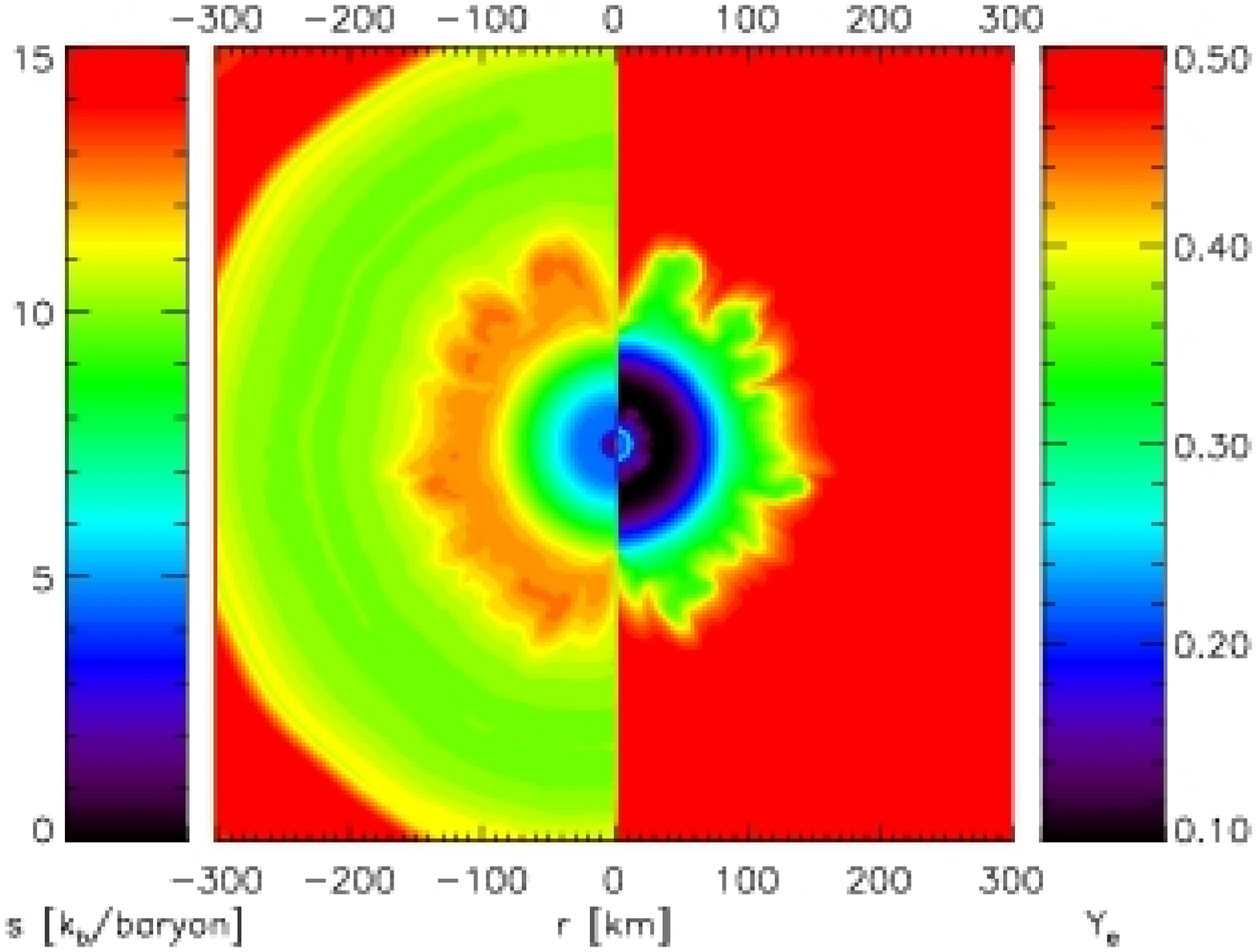} &
  \includegraphics[width=.33\textwidth]{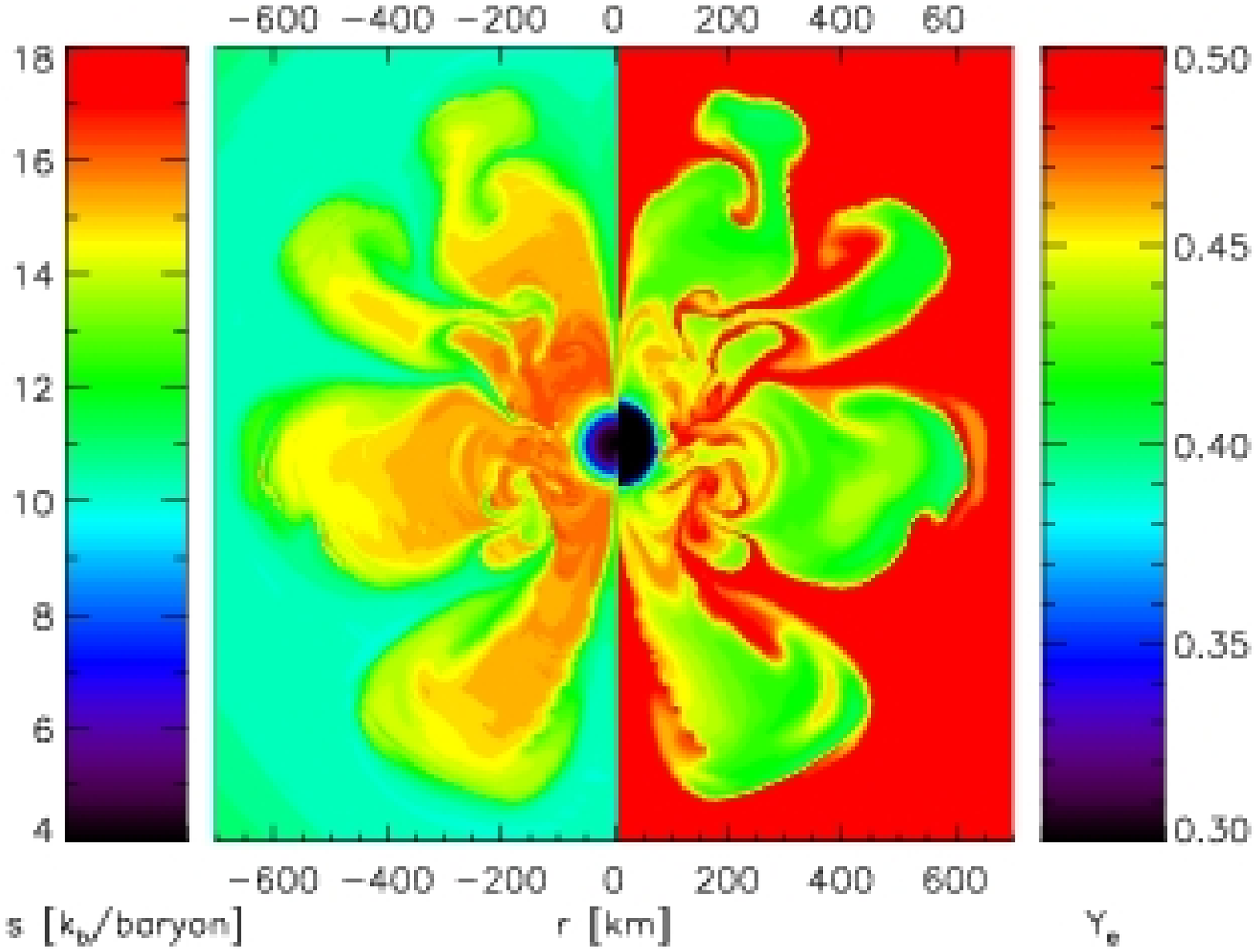} & %\\
  \includegraphics[width=.33\textwidth]{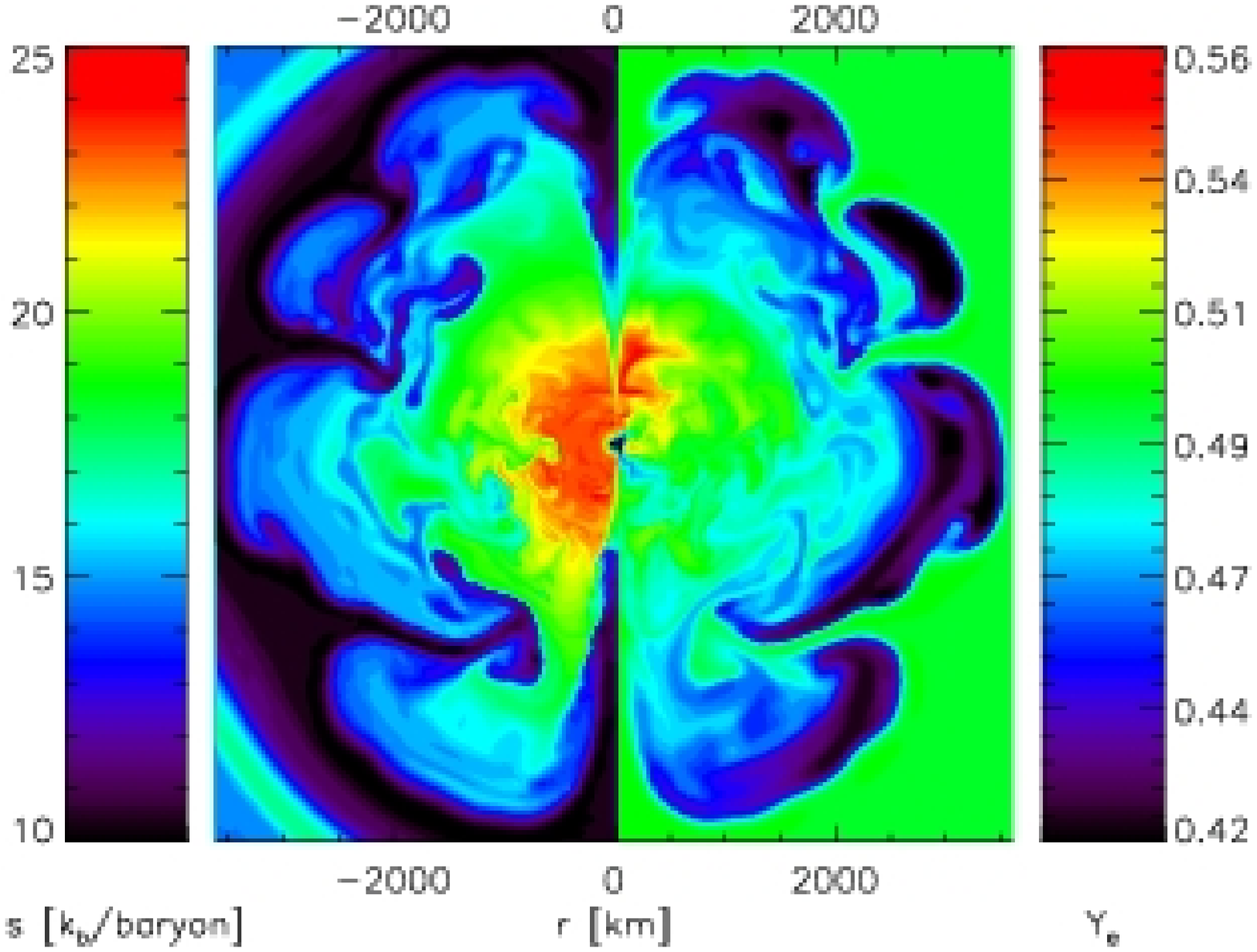}
\end{tabular}
  \caption{\label{janka_fig:snapexpl9}
Snaphots showing the gas entropy (left half of panels)
and the electron-to-nucleon ratio (right half of panels)
for the explosion of an $\sim$9$\,M_\odot$ star with
an O-Ne-Mg core.
The plots correspond (from top left to bottom right) to times of
0.097, 0.144, % 0.185, 
and 0.262 seconds after the launch of the
supernova shock front and the onset of neutron star formation at
the moment of core bounce. Note
the different radial and color scales of the four panels. Due to the
rapid expansion of the shock and of the shock-accelerated ejecta
into the extremely dilute layers surrounding the O-Ne-Mg core,
the convective pattern freezes out quickly and begins a nearly
self-similar expansion. The characteristic wavelength of convective
structures is roughly 30--45 degrees (corresponding to dominant
spherical harmonics modes of $l = 4$,$\,$5) and there is no strong
contribution of dipolar and quadrupolar asymmetries. }
\end{figure}

\begin{figure}[tpb!]
\includegraphics[width=.425\textwidth]{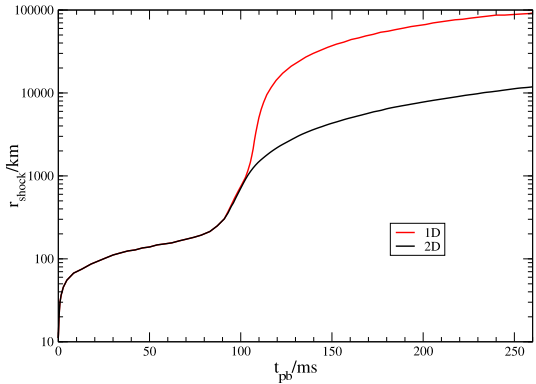} 
\caption{\label{janka_fig:shock9}
Radii of the supernova shock as functions of time
for one- and two-dimensional simulations (red and
black lines, respectively) of the explosion
of a star with O-Ne-Mg core. Note that the
progenitor used in the 2D simulation was an 8.8$\,M_\odot$ model
with an artificially constructed low-density He-shell at
$\rho < 10^3\,$g$\,$cm$^{-3}$~\cite{janka_kitaura.2006}, while the
1D simulation was performed with a recently updated progenitor
structure in which a H-envelope with a much lower density
and steeper density decline was added around the O-Ne-Mg core
(K.~Nomoto, private communication). This explains the
stronger acceleration of the shock in the region outside of
about 1100$\,$km.
}
\end{figure}

\section{Recent results}

Only in the past years the neutrino treatment in multi-dimensional
hydrodynamic and magnetohydrodynamic (MHD) models of supernovae has seen 
significant improvements. However, a 
rigorous solution of the Boltzmann transport equation is still much 
too time consuming to be applied in full-scale simulations. Even in
axially symmetric (2D) models the transport poses a five-dimensional
problem (see, e.g.,~\cite{janka_livne.2004}), in three dimensional 
hydrodynamic simulations it would constitute a time-dependent
six-dimensional problem. 

All active groups therefore still have to accept
some simplifications and nevertheless the neutrino transport module 
dominates the computing time for supernova simulations by far. The
approximations taken by different groups differ significantly. While
the Tucson-Jerusalem collaboration employs a 2D flux-limited diffusion 
scheme and treats the neutrino energy groups uncoupled (an approach
that is known from 1D simulations to be unable to capture important 
physics), thus gaining a modest amount of straightforward parallelism for 
their computations (e.g., \cite{janka_ref:burrows1,janka_ref:burrows2,janka_ref:burrows3},
the transport treatment of the Garching group accounts for the full energy
dependence of the problem and solves on each angular (lateral) bin of
the 2D grid a full one-dimensional transport problem by iterating
the moment equations of neutrino number, energy, and momentum with
a variable Eddington factor for the closure that is obtained from the 
solution of a model Boltzmann equation. Moreover, neutrino pressure
gradients and advection in the lateral direction are included in
this so-called ``ray-by-ray plus'' 
approximation~\cite{janka_ref:rampp,janka_ref:buras-I}.
While this approach assumes that the neutrino
flux components in lateral direction are zero (i.e., the neutrino 
intensity is taken to be symmetric around the radial direction), 
it allows one to properly treat the
gradual transition of neutrinos from diffusion at the high densities
in the neutron star interior to free streaming in the much more 
dilute stellar layers far outside of the neutron star. 
This approach has also the big advantage of being a
direct generalization of the 1D case and therefore it enables a 
detailed, well constrained comparison of 1D and 2D simulations.

In the following we will summarize the essentials of 
recent two-dimensional studies that have made use of the mentioned
improvements in the neutrino transport, and which have thus
contributed to a better understanding of the question how the
collapse of stellar cores could be reversed to an explosion.
The basic requirement for this to happen is that some energy
reservoir that takes up gravitational binding energy released
during stellar core collapse can be effectively tapped and 
transferred to matter that can get expelled in the explosion. 
This can happen by neutrinos in the context of the neutrino-heating
mechanism, but it can also be achieved by magnetic fields in 
magnetohydrodynamic (MHD) explosions. Or it may occur, as recently 
proposed~\cite{janka_ref:burrows1,janka_ref:burrows2}, 
through sound waves created 
by violently turbulent gas motions around the impact sites of 
accretion downflows on the neutron star surface, or even by
large-amplitude g-mode pulsations of the neutron star core, 
leading to acoustically powered explosions.

\begin{figure}[tpb!]
\tabcolsep=1mm
\begin{tabular}{lcr}
  \includegraphics[width=.33\textwidth]{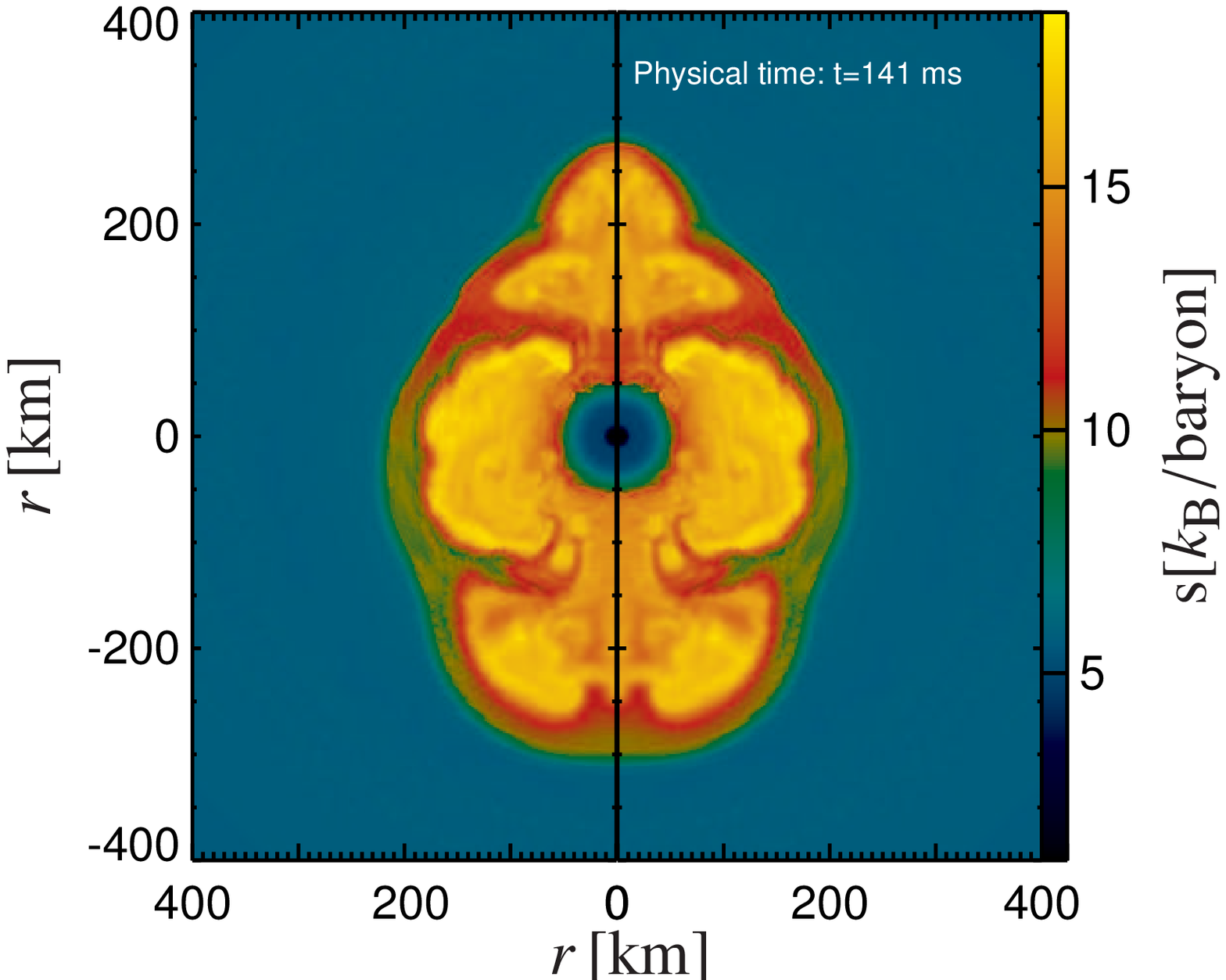} &
  \includegraphics[width=.33\textwidth]{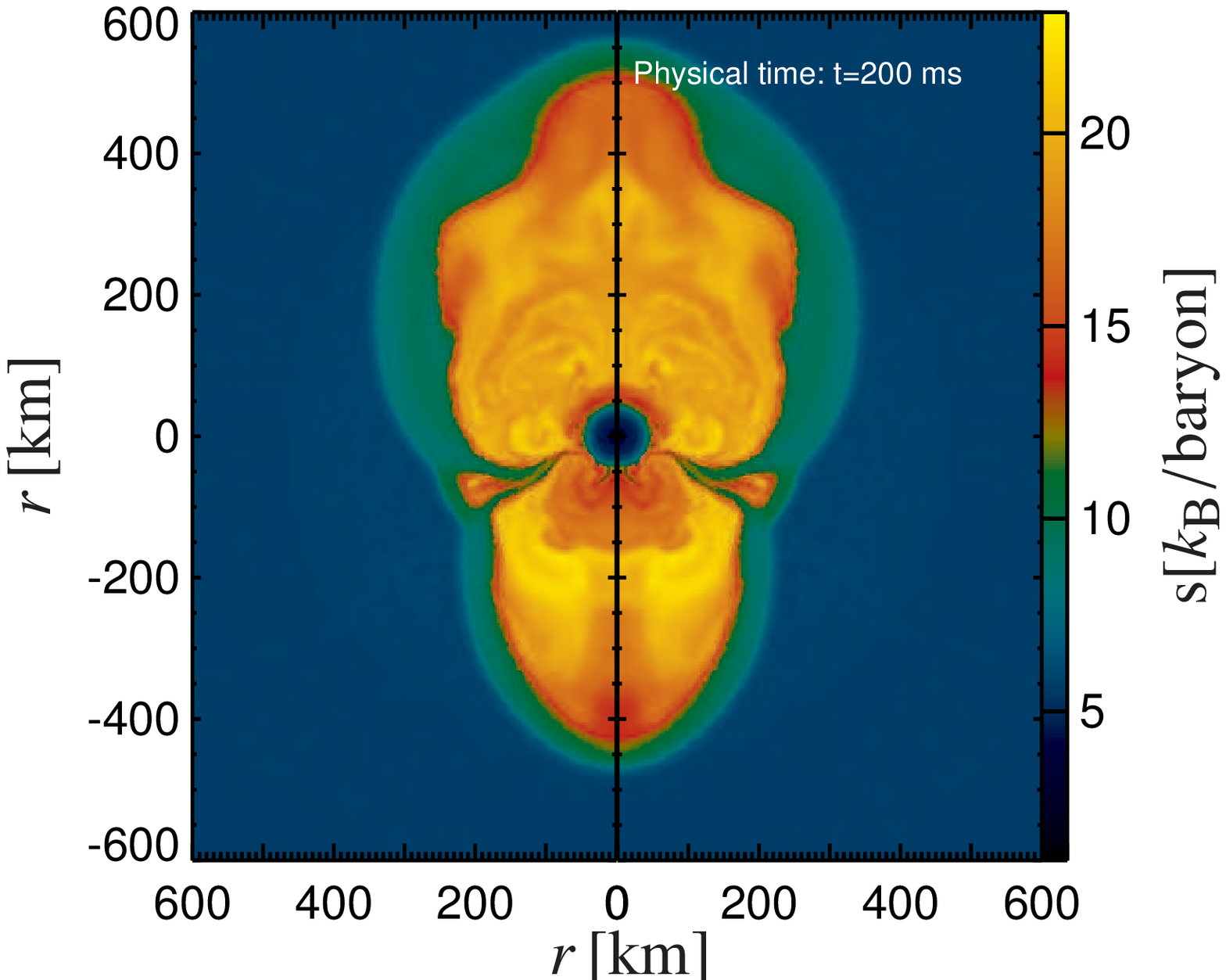} & %\\
  \includegraphics[width=.33\textwidth]{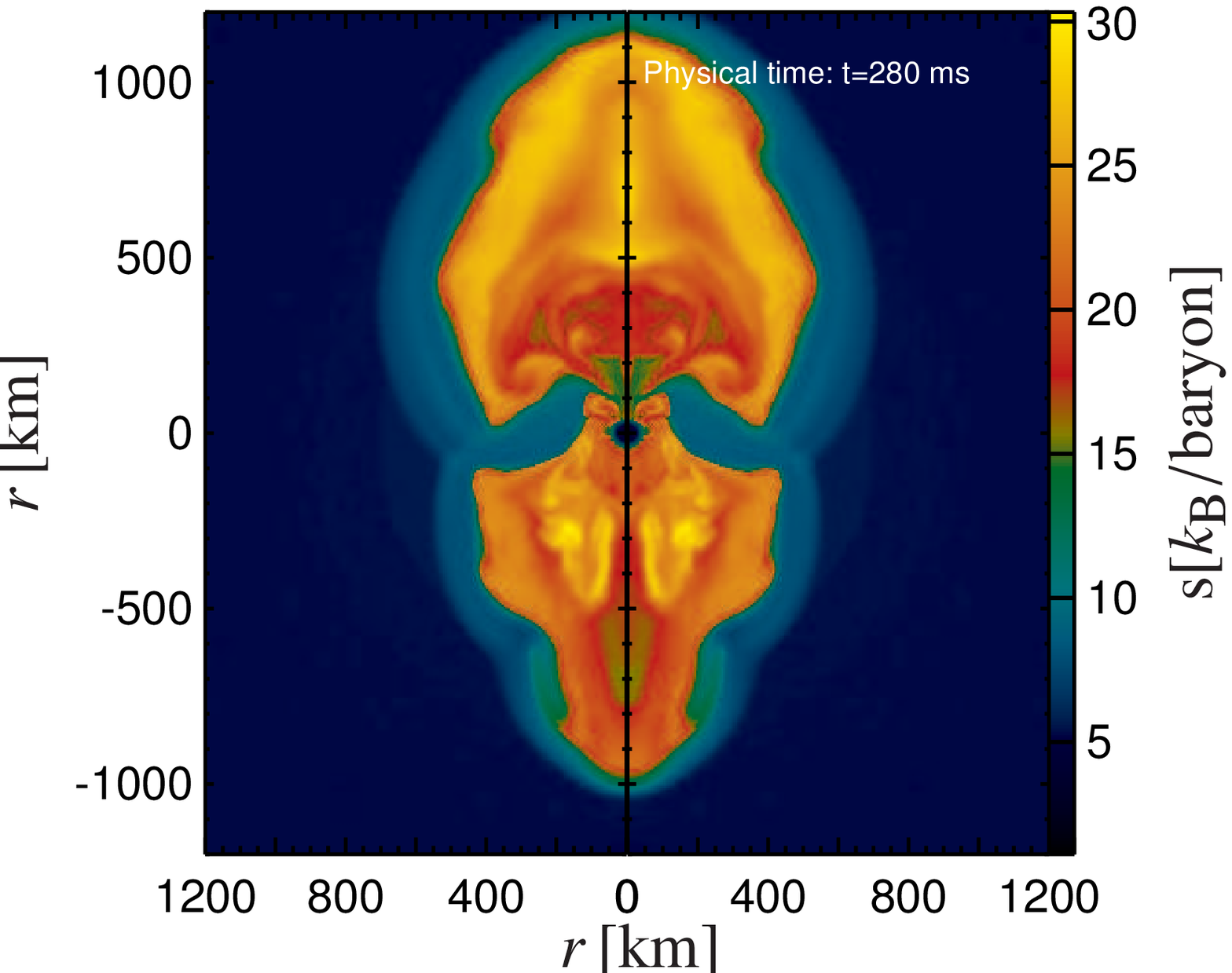}
\end{tabular}
\caption{\label{janka_fig:snapexpl11}
Snapshots of the gas entropy for the explosion of a star with
11.2$\,M_\odot$ at times 0.14, 0.20, % 0.25,
and 0.28 seconds after
the launch of the supernova shock at core bounce (top left to bottom right).
The explosion develops a large bipolar
asymmetry although the star is not rotating. Note the different radial
and entropy scales of the four panels. The first two plots show results
from Ref.~\cite{janka_ref:buras-II}, the last one is from a recent
continuation of the same simulation to later times.
}
\end{figure}

\begin{figure}[tpb!]
% \tabcolsep=3mm
% \begin{tabular}{lr}
% \includegraphics[height=.33\textwidth]{janka_fig_onemg_stoss.eps} &
\includegraphics[width=.425\textwidth]{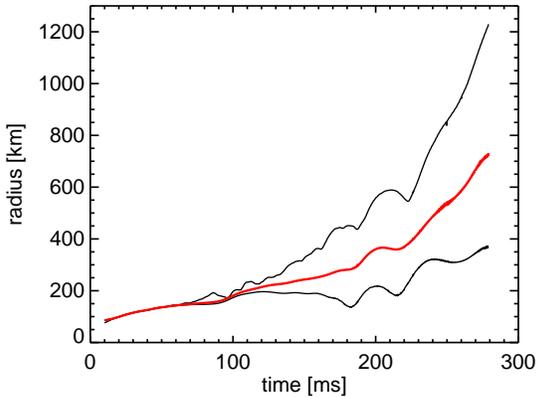}
% \end{tabular}
\caption{\label{janka_fig:shock11}
Maximum, average, and minimum radial position of the supernova
shock front as functions of time for the explosion of the
11.2$\,M_\odot$ model displayed in Fig.~\ref{janka_fig:snapexpl11}.
Note the clear signature of several
large-amplitude bipolar shock oscillations due to the standing
accretion shock instability (SASI) before
the blast takes off with an extreme 3:1 deformation.
}
\end{figure}

\subsection{Neutrino-driven explosions}

Neutrinos extract energy from the huge
reservoir of degeneracy and thermal energy that is built up
inside of the nascent neutron star during stellar core collapse.
These neutrinos diffuse out of the dense interior
and before streaming off to low-density regions, mostly neutrinos
of the electron flavor deposit roughly 10\% of their energy 
in the so-called gain layer between the gain radius and the stalled
supernova shock. Detailed and accurate numerical models are 
indispensable to determine the exact efficiency of this energy 
transfer. 

Results obtained by the Garching group for progenitor stars 
between 9 and 15 solar
masses confirm the viability of the neutrino-heating mechanism
for triggering supernova explosions. However, the inauguration of
the explosion happens in a different way than expected from 
previous models and the blast 
properties turn out to differ significantly from older
calculations with more simplified neutrino physics.

Supernova progenitors with less than about 10$\,M_\odot$ clearly
vary in their structure and explosion behavior from 
more massive stars. The former class of
stars develops a core composed of oxygen, neon, and magnesium,
not of iron, with an extremely steep density gradient at its
surface. This allows the supernova shock front, which is launched
at the moment when the neutron star begins to form at the center
of the collapsing stellar core, to expand continuously as it
propagates into rapidly diluting infalling material.
% (Fig.~\ref{janka_fig:shock+ye9}).
Behind the shock the velocities are initially negative, so that
no prompt explosion occurs. But then neutrino heating
deposits the energy that powers the ensuing blast.
Convective overturn develops in the neutrino-heated
layer behind the outgoing shock and imprints inhomogeneities on
the ejecta, in entropy as well as composition
(Fig.~\ref{janka_fig:snapexpl9}). But because of the rapid
acceleration of the supernova shock and of the postshock layer
(Fig.~\ref{janka_fig:shock9}), the pattern of
Rayleigh-Taylor structures freezes out quickly (when the 
expansion timescale becomes shorter than the overturn timescale)
and the inhomogeneous shell behind the shock begins an essentially
self-similar expansion. The short convective phase leads to 
large-scale explosion asymmetries, however, without global 
dipolar or quadrupolar deformation (Fig.~\ref{janka_fig:snapexpl9}).

Evolved stars above roughly 10$\,M_\odot$ produce
iron cores with a much more shallow density decline outside.
Running into this denser material damps the initial
expansion of the shock. Moreover, severe energy losses
and the high mass infall rates
cause the shock to even stall at a relatively small radius
of only about 100$\,$km. Because of the small shock stagnation
radius, the infall velocities of the collapsing stellar core
ahead and behind the shock are very large. Different from previous 
calculations with simple grey neutrino diffusion, our more 
sophisticated models show that convection is strongly suppressed
in the rapidly infalling matter behind the shock. Neutrino heating
is not powerful enough to allow high-entropy matter to become buoyant 
against the accretion flow 
(see~\cite{janka_foglizzo.2006,janka_ref:scheck-II}).
Convective overturn behind the shock therefore
cannot become sufficiently strong to help pushing the stagnant shock
farther out and thus to establish more favorable conditions for
neutrino heating.

Instead, another kind of nonradial hydrodynamic instability,
the so-called standing accretion shock instability 
(``SASI'';~\cite{janka_blondin.2003}),
which can grow efficiently even when convection stays
weak~\cite{janka_yamasaki.2007,janka_ohnishi.2006,janka_ref:scheck-II},
obtains decisive influence on the shock evolution. With highest
growth rates of the dipole and quadrupole modes~\cite{janka_blondin.2006}, 
it leads to violent bipolar sloshing motions of the shock. This drives 
the shock front to larger radii and thus reduces the accretion
velocities in the postshock layer. The obliqueness of the shock
surface relative to the infalling stellar core material deflects
the accretion flow and stretches its path through the layer
of neutrino heating. Moreover, the SASI also causes 
secondary convection due to steep entropy gradients produced in the
postshock layer by the quasi-periodic expansion and contraction
phases of the shock. 
The influence of the SASI thus improves the conditions for
efficient energy deposition by neutrinos, because the gas
accreted through the stalled shock can stay longer in the heating
layer and is therefore able to absorb more energy from the 
intense neutrino flux radiated by the nascent neutron 
star~\cite{janka_ref:scheck-II}.

\begin{figure}[tpb!]
\tabcolsep=1mm
\begin{tabular}{lr}
  \includegraphics[width=.550\textwidth]{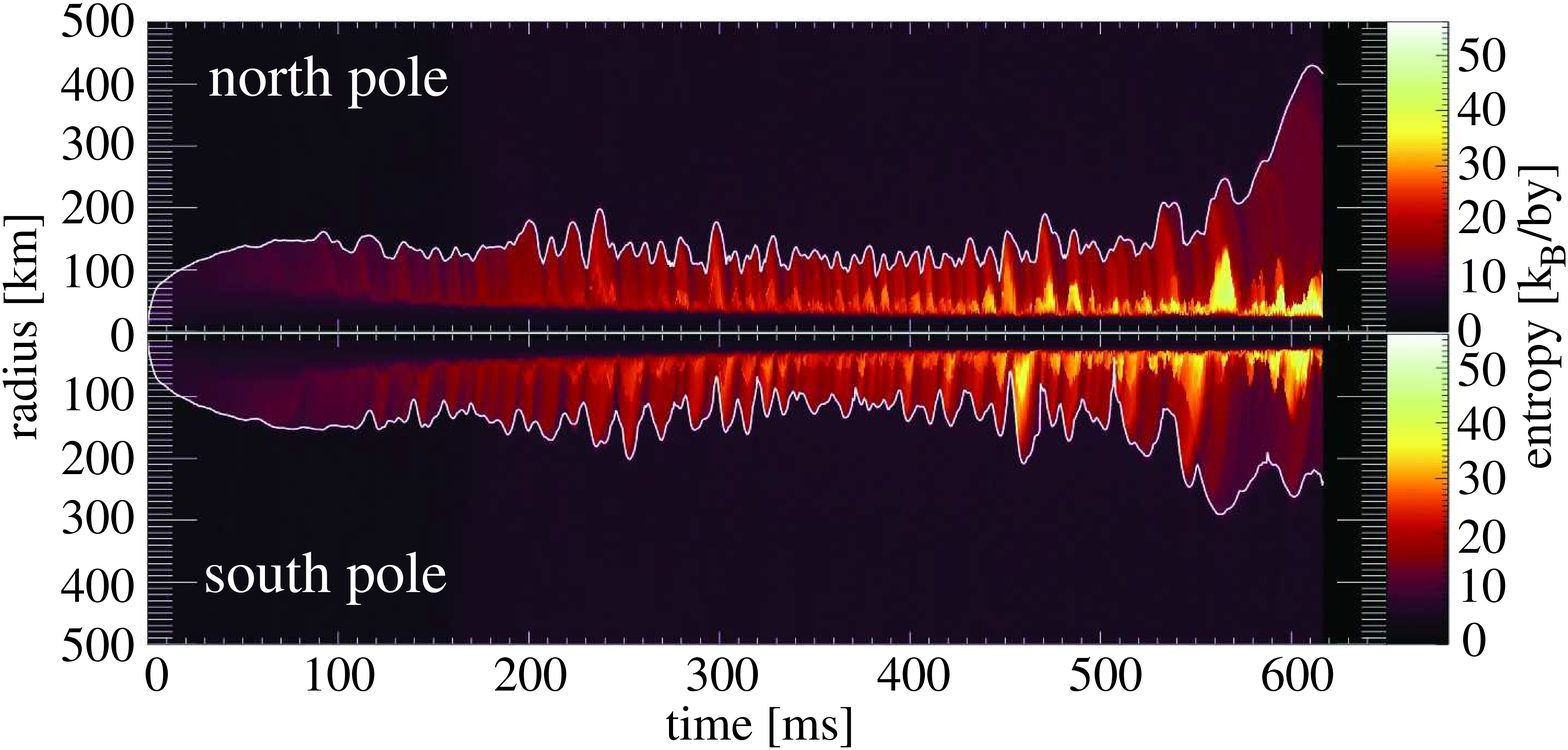} &
  \includegraphics[width=.425\textwidth]{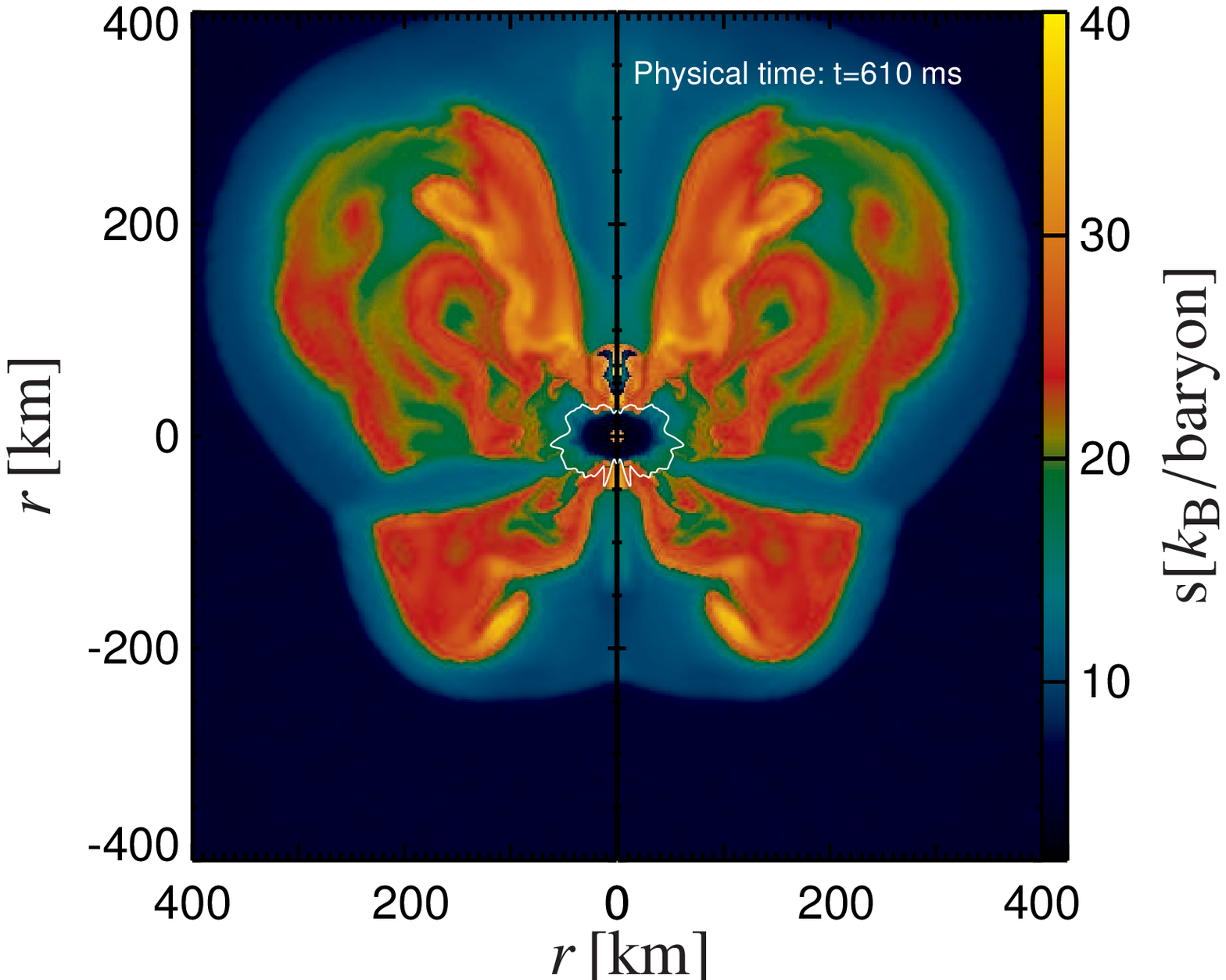} 
\end{tabular}
\caption{\label{janka_fig:expl15}
{\em Left:} Shock positions near the north pole and near the south pole
as functions of time for a 15$\,M_\odot$ star that evolves towards the
onset of an explosion at $>\,$600$\,$ms after core bounce.
The gas entropy is color coded. The plot shows many
cycles of quasi-periodic bipolar shock oscillations
due to the standing accretion shock instability (SASI). 
{\em Right:} Gas entropy distribution for the 15$\,M_\odot$
star at the onset of the explosion at a time of 0.61$\,$s after
bounce. A large north-south asymmetry signals a strong
contribution from the dipole mode. The white line marks the
gain radius as lower boundary of the neutrino-heating region
(both plots are from~\cite{janka_ref:marek}).
}
\end{figure}

The presence of strong SASI oscillations is visible in 
simulations for an 11.2$\,M_\odot$ star (for details, see
\cite{janka_ref:buras-II}) and for a 15$\,M_\odot$ star 
(details in Ref.~\cite{janka_ref:marek}). The SASI
turns out to be crucial for the explosion in both cases
(Figs.~\ref{janka_fig:snapexpl11}--\ref{janka_fig:expl15}).
Different from the $\sim$9$\,M_\odot$ star, where convection
imposes a high-mode asymmetry pattern on the ejecta
(see Fig.~\ref{janka_fig:snapexpl9}), the preferred growth
of the dipole and quadrupole ($l = 1,2$; $m = 0$ in terms of 
an expansion of the characteristic flow properties in spherical 
harmonics) 
modes of the SASI leads to a large global anisotropy of the 
beginning supernova blast in the more massive progenitors.

Our simulations have thus demonstrated the decisive role
of the standing accretion shock instability in combination
with neutrino heating for initiating neutrino-powered
supernovae. The explosion may set in
with a significant delay after the neutron star begins to form
(similar results, however with Newtonian gravity
instead of a relativistic gravitational potential were recently
reported in Ref.~\cite{janka_bruenn.2007}). 
For the most massive of the three investigated stars this
happens about 0.6 seconds later. This is not only much later than
expected from previous simulations,
but also constitutes a major computational challenge
for our 2D modeling with sophisticated multi-energy-group
neutrino transport, for which reason our set of computed cases
is still constrained to only three progenitors.

%---------------------------------------------------------------------
\begin{table}
\begin{tabular}{rrrrr}
\hline
    \tablehead{1}{r}{b}{$M_{\mathrm{prog}}$\\$[M_\odot]$}
  & \tablehead{1}{r}{b}{$t_{\mathrm{expl}}$\tablenote{Time of
   onset of explosion, determined as the moment when the total 
   energy in the gain layer, integrated over the mass elements
   with positive specific total energy,
   exceeds $10^{49}\,$erg}\\$[\mathrm{ms}]$}
  & \tablehead{1}{r}{b}{$M_{\mathrm{acc}}$\tablenote{
   Mass of accreted, neutrino-heated, and then ejected matter accounting
   for the explosion energy estimate of Eq.~(\ref{eq:expenergy})}
   \\$[M_\odot]$}
  & \tablehead{1}{r}{b}{$E_{\mathrm{expl}}$\tablenote{Explosion
    energy estimated according to Eq.~(\ref{eq:expenergy}) in
    $1\,{\mathrm{B}} = 1\,{\mathrm{bethe}} = 10^{51}\,{\mathrm{erg}}$}
    \\$[\mathrm{B}]$}
  & \tablehead{1}{r}{b}{$M_{\mathrm{ns,b}}$\tablenote{Baryonic
    mass of the neutron star at the onset of the explosion}\\$[M_\odot]$} \\
\hline
$\sim$9 & 120 & 0.01--0.02 & 0.2--0.3  & 1.36 \\
11.2    & 220 & 0.02--0.04 & 0.3--0.6  & 1.30 \\
15      & 620 & 0.05--0.06 & $\sim$1.0 & 1.55 \\
\hline
\end{tabular}
\caption{Estimated explosion properties for different progenitor stars
         with ZAMS mass $M_{\mathrm{prog}}$}
\label{tab:expprop}
\end{table}
%--------------------------------------------------------------------

Table~\ref{tab:expprop} summarizes the prediced explosion and 
remnant properties for these three progenitors. Only in the case of 
the rapidly developing blast of the $\sim$9$\,M_\odot$ star
could the 1D and 2D calculations be carried on for a sufficiently
long time to see the explosion energy asymptote. For the other two
models the explosion energy is built up during a possibly 
long-lasting phase
of anisotropic accretion and simultaneous expansion of neutrino-heated
matter, which is a manifestation of the multi-dimensional nature of the
explosion~\cite{janka_ref:scheck-I,janka_ref:burrows2,janka_ref:burrows3,janka_burrows.2007}. 
The explosion energy can then be roughly 
estimated from (for details, see~\cite{janka_ref:marek})
\begin{equation}
E_{\mathrm{exp}} \,\sim\, \dot E_\nu \,\tau_{\mathrm{acc}} 
\,\sim \,M_{\mathrm{acc}}\,e_\nu \,,
\label{eq:expenergy}
\end{equation}
where $\dot E_\nu$ is the net (i.e., heating minus cooling)
neutrino energy transfer rate to accreted and ejected gas,
\begin{equation}
\dot E_\nu  \,\sim\, \zeta \dot M_{\mathrm{acc}} \,e_\nu 
\sim 2\times 10^{51}\ \frac{\mathrm{erg}}{\mathrm{s}} \,,
\label{eq:dotenu}
\end{equation}
when $\dot M_{\mathrm{acc}}\sim 0.2$$\,M_\odot$/s is the mass 
accretion rate through the shock when the explosion begins
in the 11.2 and 15$\,M_\odot$ models, 
$\zeta\sim 0.5$ is the fraction of the accreted gas that gets
ejected again after being heated by neutrinos, 
$e_\nu \sim 10\,{\mathrm{MeV/nucleon}}\sim 10^{19}\,{\mathrm{erg/g}}$
is the average specific energy deposited by neutrinos in the gas,
and $M_{\mathrm{acc}} = \zeta \dot M_{\mathrm{acc}}\tau_{\mathrm{acc}}$
is the accreted and subsequently expelled gas mass. Accretion can continue
until the postshock matter is accelerated to escape velocity by the 
outgoing shock, which leads to an estimate of the accretion time as
$\tau_{\mathrm{acc}} \sim 0.5\,{\mathrm{s}}\,M_{\mathrm{ns},1.5} 
v_{\mathrm{sh},9}^{-3}$,  
when $M_{\mathrm{ns},1.5}$ is the neutron star mass normalized to 
1.5$\,M_\odot$
and $v_{\mathrm{sh},9}$ is the shock velocity in units of 
$10^9\,$cm/s (see ~\cite{janka_ref:marek}). 

The non-monotonic behavior of the initial neutron star (baryonic)
mass is linked to the core size of the progenitors and the 
surrounding density structure, because the latter determines
the delay of the explosion and the duration of the accretion 
phase after bounce.

\begin{figure}[tpb!]
\tabcolsep=2.5mm
\begin{tabular}{lr}
  \includegraphics[width=.425\textwidth]{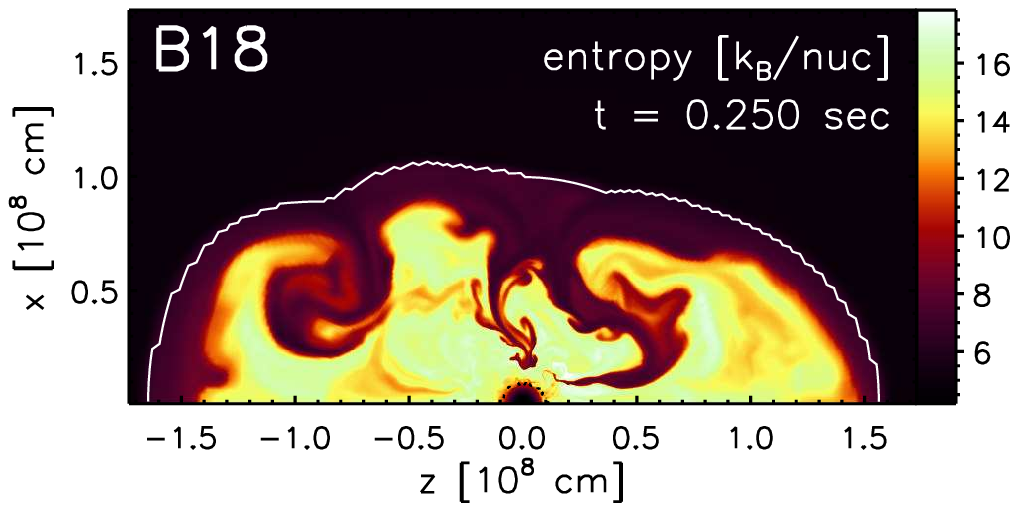} &
  \includegraphics[width=.425\textwidth]{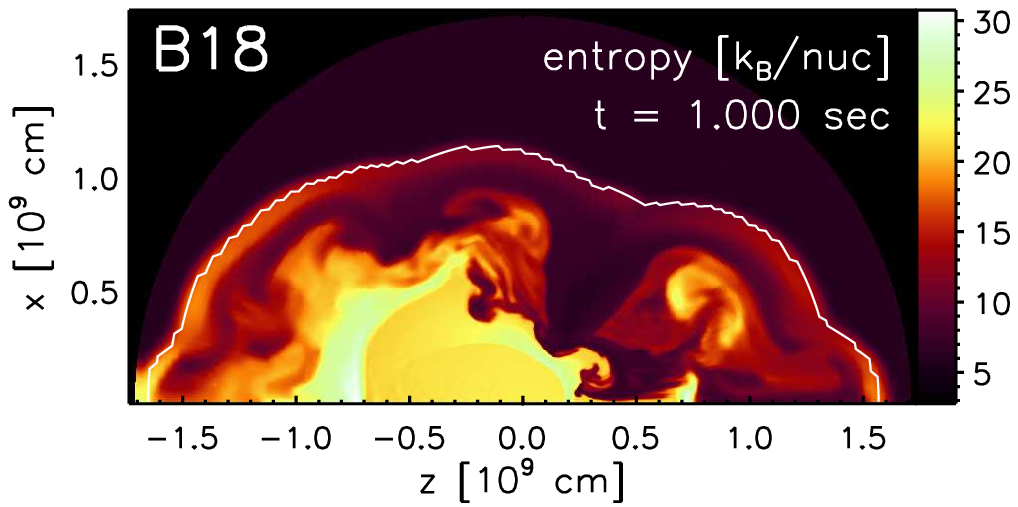}\\
  \includegraphics[width=.425\textwidth]{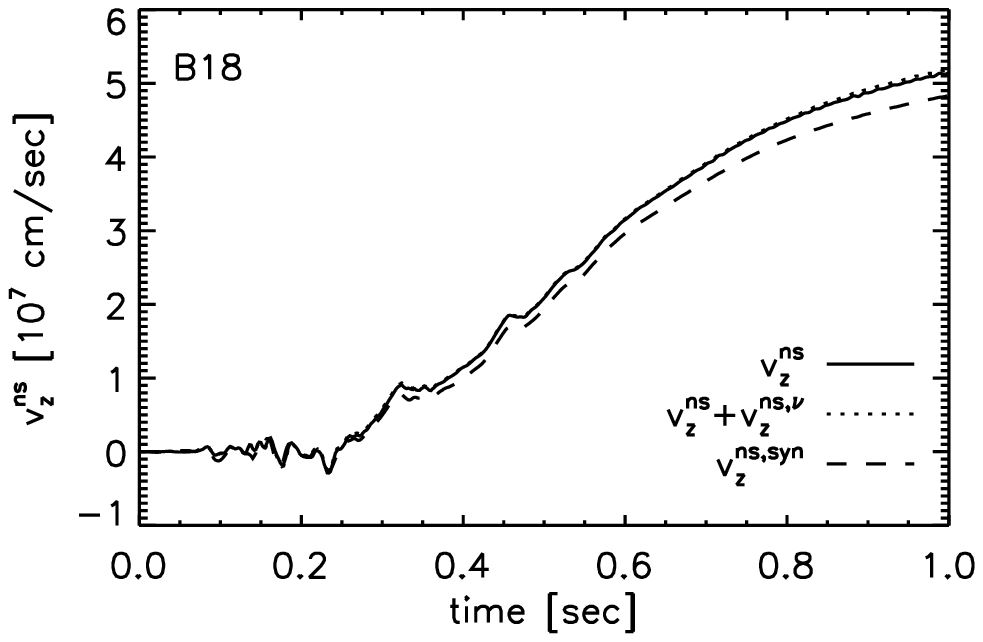} &
  \includegraphics[width=.425\textwidth]{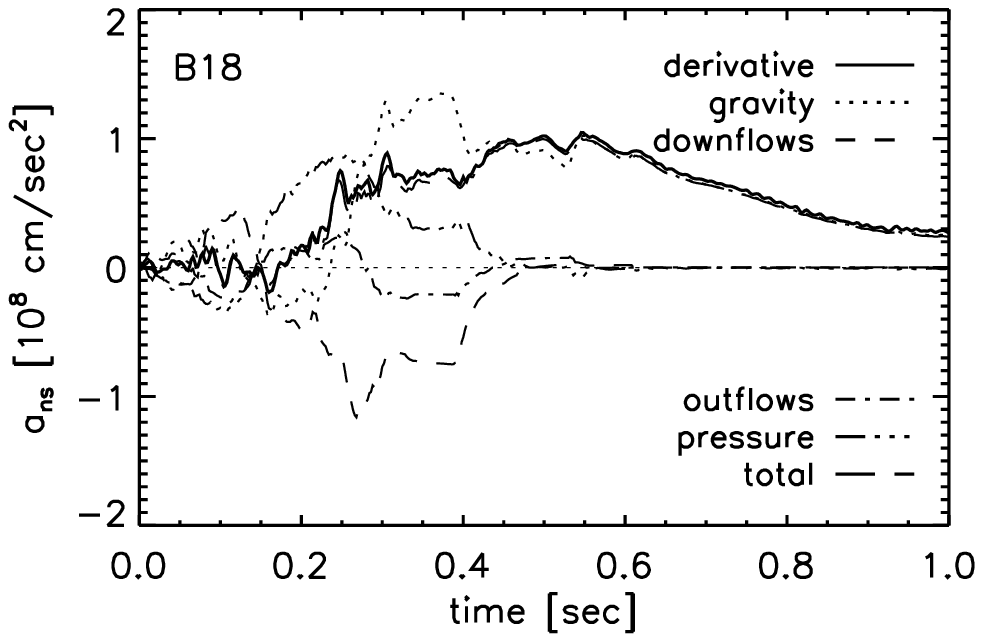}
\end{tabular}
\caption{\label{janka_fig:kick1}
{\em Upper panels:}
Two snapshots of the entropy distribution at 0.25$\,$s and 1.0$\,$s
after bounce for one of the simulations in Ref.~\cite{janka_ref:scheck-I}. 
The anisotropic ejecta distribution leads to a neutron star kick as 
explained in the text.
{\em Lower panels:}
Neutron star recoil velocity (left) and acceleration (right) as functions 
of time after bounce for the simulation shown in the upper two panels. 
The solid curves are the result deduced directly from the hydrodynamic 
simulation, the dashed curve in the left plot and the long-dashed curve 
in the right plot are from an independent post-processing analysis of
all accelerating effects on the neutron
star. It turns out that the gravitational pull of the anisotropic ejecta
is the main mediator of the neutron star acceleration 
at $t > 0.5\,$s (dotted curve in the
right panel) and dominates anisotropic accretion and outflows and pressure
forces (short-dashed, dash-dotted, and dashed-triple dotted, respectively,
in the right panel) by far. Anisotropic neutrino emission has usually a
small influence (difference between solid and dotted lines in the left
panel).
}
\end{figure}

\subsection{Explosion asymmetries and pulsar kicks}

One of the consequences of this SASI-supported neutrino-driven
mechanism is a global asymmetry of the accelerating shock front and
of the ejected gas even in the absence of rotation (or with only very
little rotation) in the stellar core. When the dipole and 
quadrupole modes are very strong, the onset of the explosion can 
resemble even a bipolar jet-like blast with a sizable pole-to-equator 
deformation (see Figs.~\ref{janka_fig:snapexpl11} and 
\ref{janka_fig:shock11}). 

A strongly deformed shock wave triggers mixing instabilities
(Rayleigh-Taylor as well as Richtmyer-Meshkov) at the
interfaces of the different composition shells of the
exploding star after the passage of the outgoing shock wave.
This was shown to lead to large-scale mixing of the chemical 
elements between the deep interior and the outer stellar layers
during the explosion, explaining self-consistently a
variety of properties observed in well-monitored supernovae
like the famous Supernova~1987A, for example the 
high nickel velocities, the inhomogeneous and clumpy distribution
of the metals, and the spreading of hydrogen over a wide range
of velocities that had to be invoked for explaining the 
smoothness and broadness of the lightcurve 
peak~\cite{janka_ref:kifonidis}. Big explosion asymmetries
can therefore not be interpreted as a signature of MHD-driven 
(``jet-driven'') explosions.

Scheck et al.~\cite{janka_ref:scheck-I}, performing a large set of
2D simulations for neutrino-driven explosions with an approximative
description of the neutrino transport and using
the neutrino luminosities from the contracting and cooling neutron 
star as a parametric boundary condition, showed that such 
large explosion asymmetries can leave the compact remnant
with recoil velocities sufficiently large to explain the measured
eigenvelocities of young pulsars. In cases where a dipolar 
asymmetry became dominant and the explosion developed more
strength in one hemisphere than in the other, typical kick
velocities around 500$\,$km/s were found, with peak values
even above 1000$\,$km/s, whereas in cases where the higher modes 
were stronger than the $l=1$ asymmetry, the velocities stayed 
fairly modest, usually below about 200$\,$km/s 
(see Fig.~20 in~\cite{janka_ref:scheck-I}).

The pulsar recoil is caused by the asymmetry which the 
SASI distorted explosion develops on the long run, i.e. over
a timescale of many seconds. During the
onset of the explosion, even until the shock reaches a radius
of some 1000$\,$km, the pulsar kick usually does not grow to
large values, which indicates that the ejecta have not obtained
a high momentum asymmetry until then. 
In Fig.~\ref{janka_fig:kick1}, which displays one
of the cases computed in Ref.~\cite{janka_ref:scheck-I},
the left panels show that the neutron star is essentially
not accelerated until 250$\,$ms after bounce when the maximum
shock radius in this simulation is beyond 1500$\,$km, 
and even at 400$\,$ms
post bounce the neutron star has attained a velocity of less
than 100$\,$km/s. Only later the acceleration grows and 
leads to a recoil velocity that asymptotes much after one
second post bounce. The reason for such a long-lasting neutron star
propulsion can neither be anisotropic accretion nor anisotropic
mass ejection in the neutrino-driven wind. The former ceases at
about 0.5$\,$s p.b., while the neutrino wind is essentially
spherically symmetric, corresponding to a neutrino emission that
is nearly isotropic and thus also produces only a very small
effect on the neutron star kick velocity (see left lower panel
of Fig.~\ref{janka_fig:kick1}). 

A careful analysis of all 
effects that can transfer momentum between the surrounding gas
and the compact remnant, i.e., gas outflow and accetion, anisotropic
pressure, neutrino emission, and gravitational forces, shows that
mostly the last mediate the speed-up of the neutron star at
$t > 0.4\,$s, whereas before that time all individual forces are 
large but nearly balanced
(see lower right panel of Fig.~\ref{janka_fig:kick1}
and for details, Ref.~\cite{janka_ref:scheck-I}). The explosion 
asymmetries, which are the cause of the gravitational momentum 
transfer to the neutron star, develop only gradually when the
outward moving shock encompasses more and more matter
from the progenitor star. Since the shock was launched highly 
aspherically as a result of the SASI motions, the gas swept
up by the shock may not experience the same acceleration 
everywhere. If the shock is weaker in a certain direction or
is oblique to the radius vector, the swept up gas is less
strongly accelerated. This slower gas begins to lag behind the 
ejecta expanding in other directions. It is funneled into dense,
downward-reaching, low-entropy filaments with the typical
mushroom-like Rayleigh-Taylor caps (these are 
visible in the right upper panel of Fig.~\ref{janka_fig:kick1}).
The gravitational attraction of such massive gas pockets, which
are closer to the neutron star than the ejecta in other directions,
excerts an anisotropic pull on the compact remnant. Reversely, this
pull decelerates the still expanding gas, thus transferring some
of the ejecta momentum to the neutron star. In the extreme case,
the gas may be gravitationally captured and may fall back to be
accreted by the neutron star. The acceleration ceases and the 
neutron star speed asymptotes, when the inhomogeneous 
ejecta reach larger
and larger radii and the anisotropic gravitational forces 
diminish. As a consequence of this acceleration, the neutron
star motion is predicted to be opposite to the main momentum
of the matter ejected in the supernova blast.

\begin{figure}[tpb!]
\tabcolsep=0mm
\begin{tabular}{c}
  \includegraphics[width=.475\textwidth]{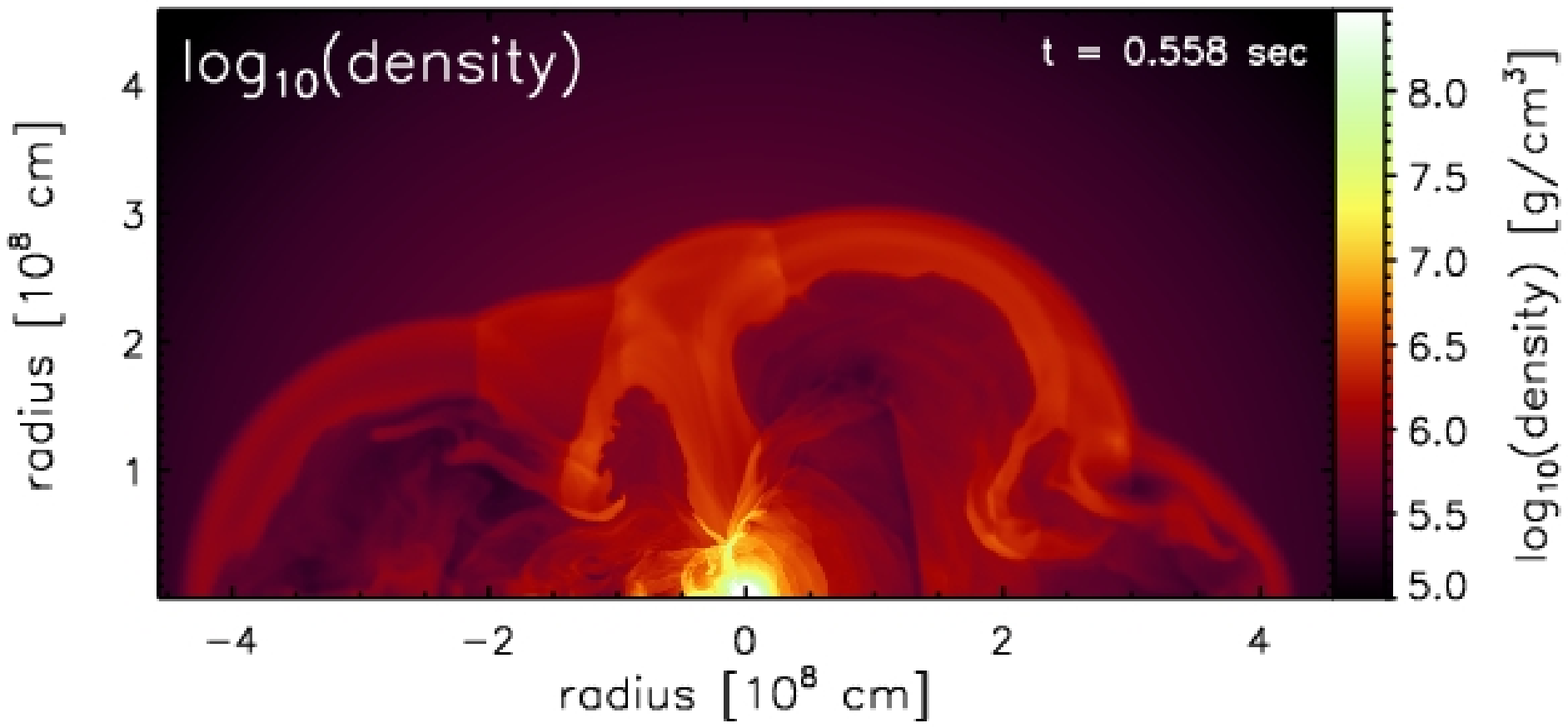}\\ %& 
   \includegraphics[width=.475\textwidth]{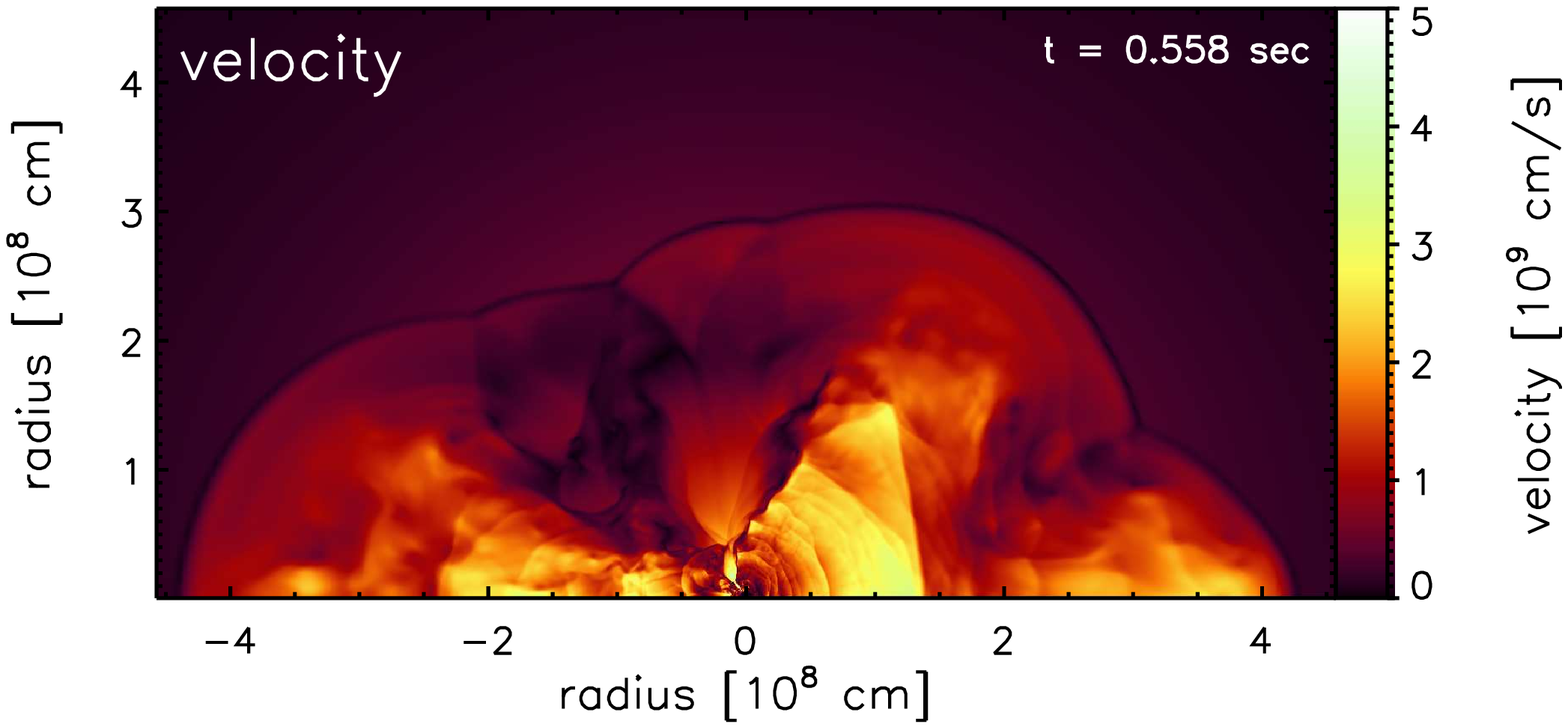}\\ %&
  \includegraphics[width=.475\textwidth]{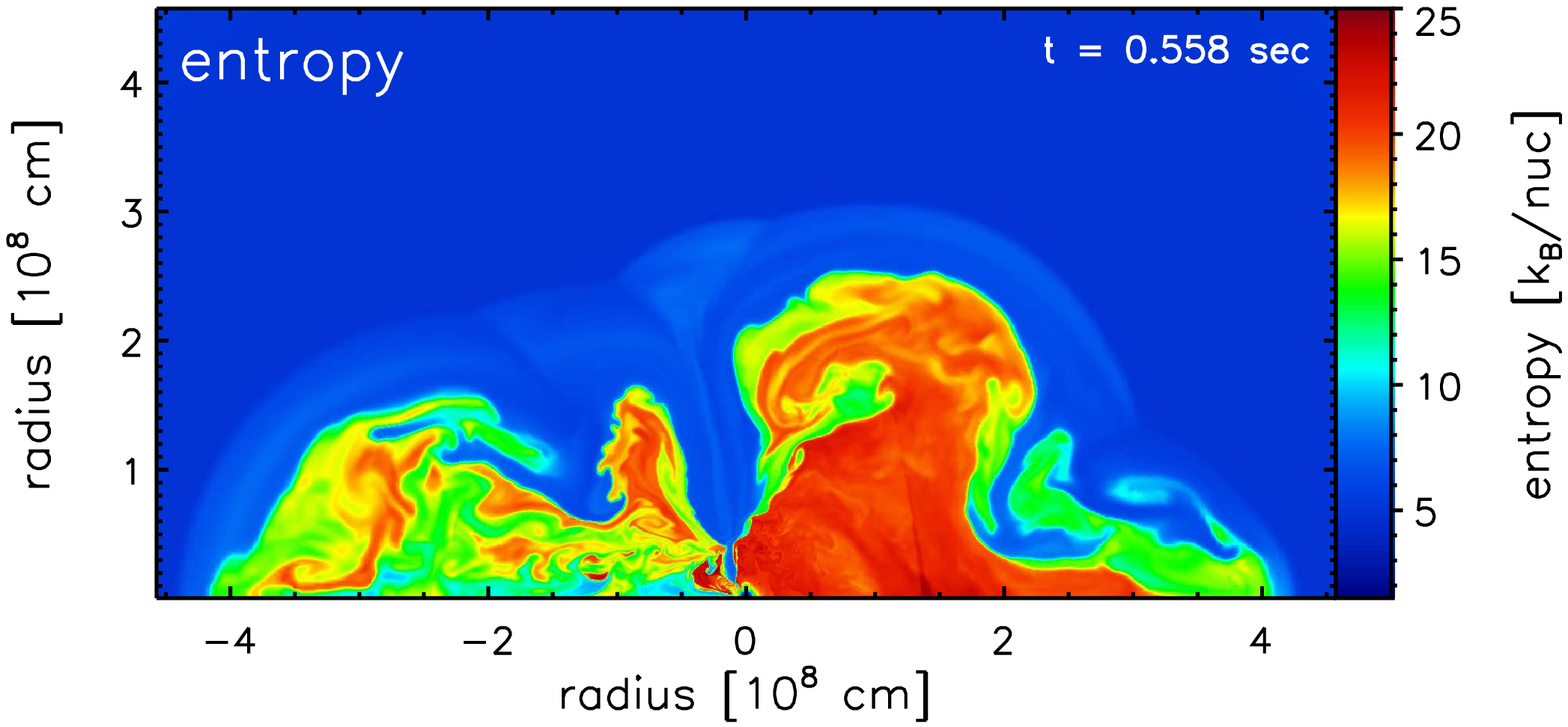} %&
\end{tabular}
\caption{\label{janka_fig:waves1}
Snapshots of density, absolute value of the radial velocity, and entropy
per nucleon (from top to bottom)
at $\sim$0.56$\,$s after supernova shock formation in one of the 
explosion simulations of a 15$\,M_\odot$ studied in 
Ref.~\cite{janka_ref:scheck-I}. The anisotropic density distribution 
of the ejecta is visible in the upper plot, with mushroom-like
Rayleigh-Taylor structures reaching down towards the neutron star
at the grid center. The middle panel shows the absolute values 
of the radial velocity, which reveals strong sonic activity
due to the impact of the downflows on the neutron star surface, and
the lower plot demonstrates that the outgoing sound waves do not dissipate
their energy efficiently in the rapidly expanding neutrino-heated
gas so that the entropies of these ejecta are hardly affected.
}
\end{figure}

\begin{figure}[tpb!]
\tabcolsep=0.5mm
\begin{tabular}{lr}
  \includegraphics[width=.235\textwidth]{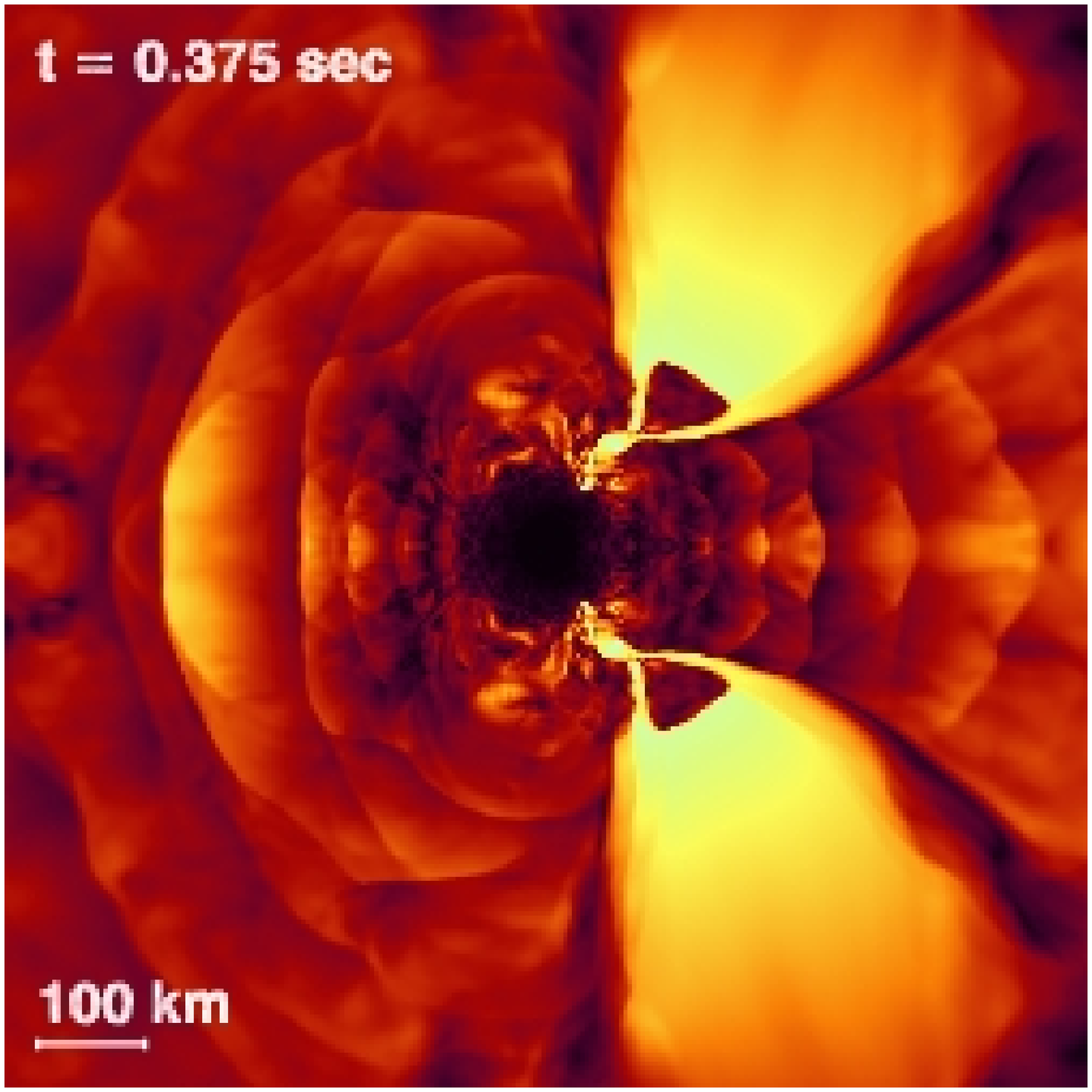} &
  \includegraphics[width=.235\textwidth]{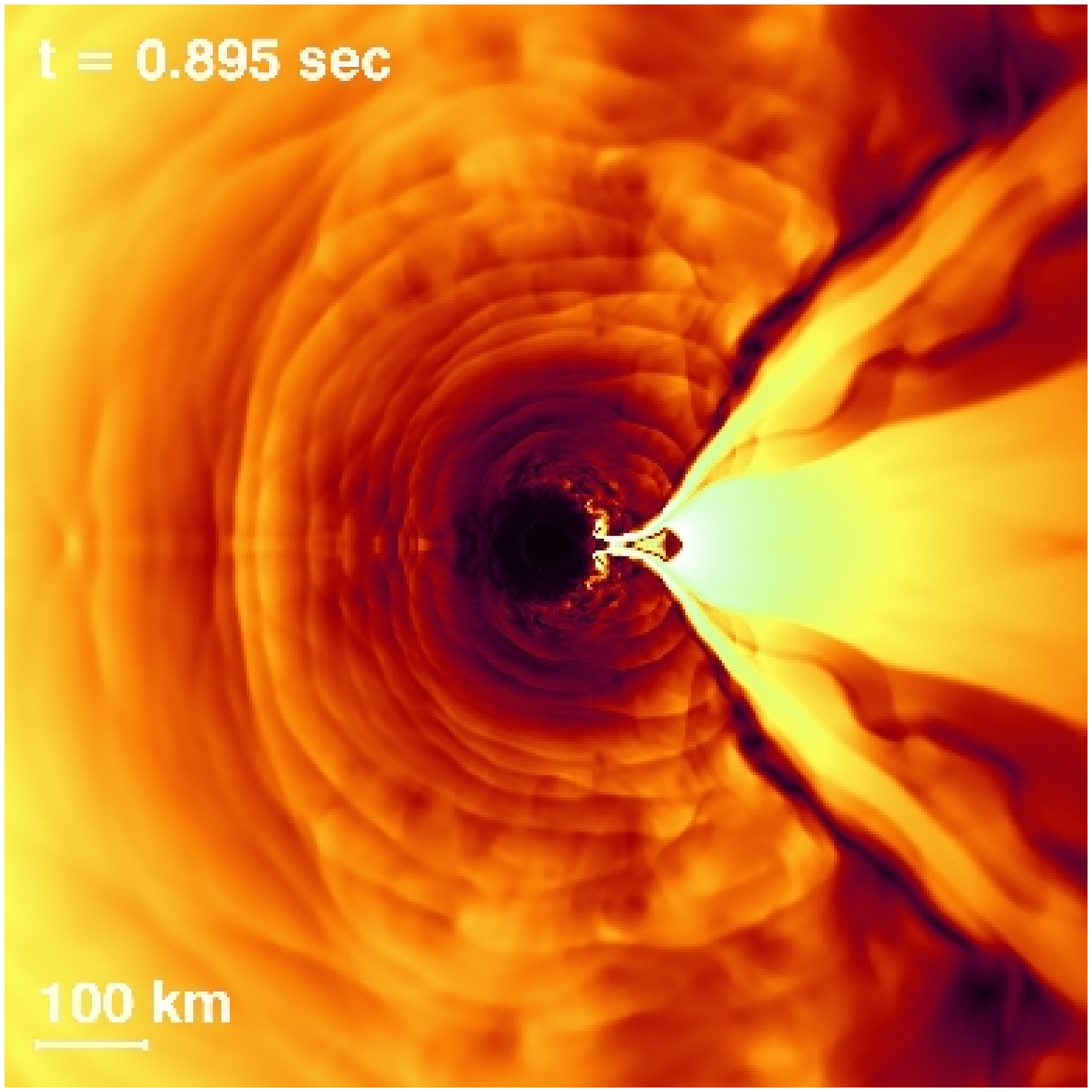} 
\end{tabular}
\caption{\label{janka_fig:waves2}
Close-up of the neutron star vicinity for two post-bounce times
(0.375$\,$s and 0.895$\,$s p.b.) in a simulation like the one 
shown in Fig.~\ref{janka_fig:waves1}. The color coding represents the 
absolute value of the radial velocity as in the middle panel of
the latter figure. The sonic activity by waves emerging from the 
impact of the accretion flow on the neutron star surface can be 
particularly well seen in this quantity.
}
\end{figure}

\subsection{Alternative explosion mechanisms}

Recently, Burrows et al.~\cite{janka_ref:burrows1,janka_ref:burrows2}
came up with the suggestion that supernovae might be energized
by a strong flux of acoustic power originating from the neutron star.
In their 2D simulations they found that the compact remnant is 
instigated to large-amplitude bipolar oscillations, $l=1$ core
gravity modes, by the anisotropic accretion of gas. The pulsating
compact remnant sends pressure waves into its environment,
which carry a sizable energy flux and can even steepen into
shocks, thus dissipating their energy in the surrounding medium
and raising the entropy there. The ringing neutron star acts
as a transducer that converts some part of the gravitational 
binding energy released by the accreted gas into sonic power.
The radiated sound was found to be the crucial supply of the
developing explosion with energy and momentum and thus triggers
acoustically driven explosions.

While this appears as an interesting alternative to initiate
the explosion if neutrino heating fails, the question is how
the acoustic energy input compares to neutrino heating.
Burrows et al.~\cite{janka_ref:burrows1} reported that in their 
simulations the acoustic energy flux dominates the neutrino
energy deposition later than several 100$\,$ms after bounce.
Although we do not observe the large $l=1$ core g-modes in the neutron
star seen by them at late post-bounce times, and on the basis
of our present simulations we can neither judge nor exclude this 
possibility, 
also in our models ---in the full-scale supernova calculations
for $\sim$8--15$\,M_\odot$ stars as well as in the parametric
explosions--- we can clearly identify the presence of strong 
turbulence in the neutron star surface layer.
The gas there is stirred by the violent impact of 
accretion downflows, thus creating vigorous sonic activity
around the neutron star (Figs.~\ref{janka_fig:waves1} and
\ref{janka_fig:waves2}). Very approximately, we can estimate
the outgoing flux of sonic energy (making the assumptions of
spherical symmetry and negligible energy dissipation in the
flow) as~\cite{janka_landau.1991} 
\begin{eqnarray}
\dot E_{\mathrm{sound}} &\sim & 4\pi r^2 \rho v^2 c_{\mathrm{s}}\,
\sim\, 4\pi r^2 \left(\frac{\rho'}{\rho}\right)^{\! 2} \rho 
c_{\mathrm{s}}^3 \nonumber \\
&\sim& 5\times 10^{50}\ \frac{\mathrm{erg}}{\mathrm{s}}\,
\left(\frac{\rho'}{\rho}\right)^{\! 2}\ ,
%\left(\frac{r}{300\,\mathrm{km}}\right)
%\rho_7\,c_{\mathrm{s},1.7}
\label{eq:esound}
\end{eqnarray}
where $v$ is the fluid velocity in the sound waves,
$c_{\mathrm{s}}$ is the sound speed, and $\rho'/\rho$ denotes
the amplitude of the ripples on the background density $\rho$ 
caused by the sound waves. The numerical value in the last
expression turns out to be fairly independent
of radius and time at some distance from the neutron star in 
Figs.~\ref{janka_fig:waves1} and \ref{janka_fig:waves2}.
For $\rho'/\rho$ of order unity our estimate agrees with 
values quoted in Burrows et al.~\cite{janka_ref:burrows2},
which might account for a sizable contribution to the energy
of the developing explosion.
In our simulations, we indeed see that the perturbations can
reach such amplitudes, in particular when the downflow becomes
unstable and the perturbations it generates in and around
the neutron star surface layer
grow. Outside of these transient periods, however, $\rho'/\rho$ is 
usually significantly smaller. Moreover, the net 
neutrino energy deposition rates, integrated over
the volume of the gain layer, 
in all of our simulations are typically several $10^{51}\,$erg/s
(see Eq.~\ref{eq:dotenu}), 
so that the energy input to the explosions in our models is
clearly dominated by neutrinos.

Magnetohydrodynamically driven
explosions have attracted renewed and increasing interest
over the past years for a number of reasons 
%(see, e.g., \cite{janka_akiyama.2003,janka_thompson.2005,janka_takiwaki.2007,janka_obergaulinger.2006,janka_moiseenko.2007,janka_ref:burrows3}).
One reason is the discovery
of magnetars, which is taken as an indication that at least in
some fraction of all cases very strong surface magnetic fields
can be generated in neutron stars, possibly already during the
evolution phases in which a supernova explosion is triggered.
Another reason is the established link of at least some long-duration
gamma-ray bursts 
with very energetic and highly aymmetric hypernova explosions of
presumably rapidly rotating massive stars. And a third reason
is the interpretation of asymmetries observed in supernova
explosions and supernova remnants as consequences and relics of
jet-like eruptions or as hints to the ejection of highly collimated
material.

Independent of how magnetic fields transfer energy to the 
explosion, e.g., by hoop stresses, magnetic pressure, or viscous
dissipation of energy, the initial magnetic fields in the stellar
core must have been amplified during the collapse either by magnetic
field wrapping or the magnetorotational instability. Thus they tap
the free energy that can be stored in the highly differential rotation 
of a spinning stellar core that collapses at angular momentum conserving
conditions. This free energy, however, is typically a rather small
fraction (of order 10\%) of the total rotation energy, i.e.,
\begin{equation}
E_{\mathrm{rot}}^{\mathrm{free}}\,<\,
E_{\mathrm{rot}}\,\approx\, 2\times 10^{52}\,{\mathrm{erg}}\
M_{\mathrm{ns},1.5} R_{\mathrm{ns},6}^2 
\left({1\,{\mathrm{ms}}\over P_{\mathrm{ns}}}\right)^{\! 2}\,,
\end{equation}
where $R_{\mathrm{ns},6}$ denotes the neutron star radius in 
$10^6\,$cm and $P_{\mathrm{ns}}$ is the neutron star spin period.
A millisecond neutron star typically requires the pre-collapse
stellar core to rotate with a period of $P_{\mathrm{ini}}\sim
P_{\mathrm{ns}}(R_{\mathrm{ini}}/R_{\mathrm{ns}})^2\sim 
10\,{\mathrm{s}}\,(P_{\mathrm{ns}}/1\,{\mathrm{ms}})$. This is
significantly faster than predicted by current evolution models for
rotating stars, in which the enhanced angular momentum loss 
due to angular momentum transport by magnetic field
effects is taken into account~\cite{janka_heger.2005}.
At the onset of core collapse,
the stellar cores are estimated to have spin periods of 
$P_{\mathrm{ini}} > 100\,$s (i.e., $\Omega_{\mathrm{ini}} < 0.05\,$rad/s).
This will lead to neutron stars with $P_{\mathrm{ns}} > 10\,$ms,
which is much too slow for rotation to be an energy reservoir of
MHD-driven supernovae. Such a conclusion was also reached on the
basis of detailed simulations in 
Refs.~\cite{janka_thompson.2005,janka_ref:burrows3}. 
Nevertheless, the
MHD mechansim may still need to be invoked for explaining the 
enormous energy output of long-duration gamma-ray bursts and
associated hypernova explosions, which would imply that these
events are linked to rare cases
where massive stars have achieved to retain a large angular 
momentum at the time when they end their lives as 
collapsars~\cite{janka_macfadyen.1999}.

\section{Summary and outlook}

We have reviewed a variety of recent findings that shed
new light on the processes that cause the explosions of 
massive stars and play a role during the birth of neutron stars. 
All recent 2D simulations that were performed with a
full 180 degree grid agree that the 
standing accretion shock becomes unstable to low-mode, nonradial
deformation. This SASI phenomenon plays a very important role
in the supernova core. It was not only found to induce a large
asymmetry of the developing blast but also to 
facilitate neutrino-driven explosions by stretching the time
accreted matter can stay in the gain layer and can be exposed to
neutrino heating. Moreover, the SASI was observed to excite 
large-amplitude core g-modes in the neutron star, whose sonic
damping could contribute to or be essential for powering the
explosion. The asymmetries imprinted on the explosion by the
SASI may lateron lead to neutron star kicks and might explain
the observed velocities of young pulsars. In three dimensions
the $m\neq 0$ modes of the SASI can also have an influence on
the spin of the forming neutron star~\cite{janka_blondin.2007}.

Despite the general agreement about the importance of the SASI,
the physical mechanism behind this phenomenon is still a matter
of vivid debate (see~\cite{janka_ref:scheck-II} and references therein)
and the present calculations differ in many conclusions. Partly
this is may be so because of
the significantly different numerical approaches taken,
e.g., regarding the hydrodynamics and neutrino transport, 
the computational grid, the description of gravity, and
the employed equation of state and neutrino interactions.
Some of the current 
discrepancies and controversies, however, will find an 
explanation when more simulations become available and 
comparisons are made. Ultimately, however, most of the 
processes and consequences mentioned above will have to 
be addressed by 3D models, which are currently still out of
reach because of the enormous computational demands of the
energy-dependent neutrino transport.

\begin{theacknowledgments}
We thank K.~Nomoto, A.~Heger and S.~Woosley for
providing us with their progenitor data
and are indebted to R.~Buras, K.~Kifonidis, E.~M\"uller, and
M.~Rampp for fruitful collaborations.
This project was supported by the Deutsche Forschungsgemeinschaft
through the Transregional Collaborative Research Centers SFB/TR~27
``Neutrinos and Beyond'' and SFB/TR~7 ``Gravitational Wave Astronomy'',
%  the Collaborative Research Center SFB-375 ``Astro-Particle Physics'',
and the Cluster of Excellence ``Origin and Structure of the Universe''
(\url{http://www.universe-cluster.de}). The computations were only
possible because of computer time on the IBM p690 of
the John von Neumann Institute for Computing (NIC) in J\"ulich,
on the national supercomputer NEC SX-8
at the High Performance Computing Center Stuttgart (HLRS) under
grant number SuperN/12758, on the IBM p690 of the Computer Center
Garching (RZG), on the sgi Altix 4700 of the Leibniz-Rechenzentrum (LRZ)
in Munich, and on the sgi Altix 3700 of the MPI for Astrophysics.
We also acknowledge support by AstroGrid-D, a project funded by
the German Federal Ministry of Education and Research (BMBF)
as part of the D-Grid initiative.
\end{theacknowledgments}

\end{document}